Full paper

# Employing surfactant-assisted hydrothermal synthesis to control CuGaO$_2$ nanoparticle formation and improved carrier selectivity of perovskite solar cells

Ioannis T. Papadas[a], Achilleas Savva[a], Apostolos Ioakeimidis[a], Polyvios Eleftheriou[a], Gerasimos S. Armatas[b] and Stelios A. Choulis[a]*

[a] *Molecular Electronics and Photonics Research Unit, Department of Mechanical Engineering and Materials Science and Engineering, Cyprus University of Technology, Limassol, Cyprus.*

[b] *Department of Materials Science and Technology, University of Crete, Heraklion 71003, Greece.*

*Corresponding Author: Prof. Stelios A. Choulis

E-mail: stelios.choulis@cut.ac.cy

ABSTRACT

Delafossites like CuGaO$_2$ have appeared as promising p-type semiconductor materials for opto-electronic applications mainly due to their high optical transparency and electrical conductivity. However, existing synthetic efforts usually result in particles with large diameter limiting their performance relevant to functional electronic applications. In this article, we report a novel surfactant-assisted hydrothermal synthesis method, which allows the development of ultrafine (~5 nm) monodispersed p-type $CuGaO_2$ nanoparticles (NPs). We show that DMSO can be used as a ligand and dispersing solvent for stabilizing the $CuGaO_2$ NPs. The resulting dispersion is used for the fabrication of dense, compact functional $CuGaO_2$ electronic layer with properties relevant to advanced optoelectronic applications. As a proof of concept, the surfactant-assisted hydrothermal synthesized $CuGaO_2$ is incorporated as a hole transporting layer (HTL) in the inverted p-i-n perovskite solar cell device architecture providing improved hole carrier selectivity and power conversion efficiency compared to conventional PEDOT:PSS HTL based perovskite solar cells.

**1. Introduction**

The extraordinary characteristics of organic-inorganic lead halide perovskites, such as high light absorption [1-4] enhanced charge transport properties and direct band gap transition have improved the photovoltaic performance rapidly during the last years. As a result, a power conversion efficiencies (PCE) surpassing 20 % have been reported [5-9]. The so-called p-i-n perovskite architecture is widely applied for the fabrication of efficient perovskite solar cells [10,11]. In this structure, poly(3,4-ethylenedioxythiophene): poly(styrenesulfonate) (PEDOT:PSS) is the most commonly used hole transporting layer (HTL) due to its facile processing, good electrical conductivity and transparency [11-15]. Nevertheless, PEDOT: PSS layers might lead to carrier selectivity limitations due to its energy levels, and high hygroscopicity [16,17].

The implementation of p-type metal oxides as HTLs, such as NiOx and CuOx [18-20], in photovoltaic devices can enhance the overall performance of the cell due to improved hole selectivity and chemical stability [21,22]. Despite these essential characteristics, the low conductivity and high absorption coefficient induce the need for very thin films which usually leads to low reproducibility of high PCE photovoltaic devices [18-20]. Therefore, the fabrication of alternative p-type HTL with improved carrier selectivity is needed [23].

Delafossites is one of the few categories of p-type transparent conductive oxides (TCOs) that combine high electrical conductivity and optical transparency [23-25]. Thus, delafossites have been proposed as functional materials for various advanced electronic applications, including p–n junction transparent electronics, photo and electrochemical catalysis, such as water splitting and reduction of $CO_2$, and different types of solar cells [23-34].

Specifically, the delafossite $CuGaO_2$ shows a wide optical band gap in the range of 3.4–3.7 eV [35], low valence band (VB) edge position (~5.4 eV vacuum level), high hole mobility and exceptional chemical and thermal stability [30,36-38]. However, the high preparation temperature required for the formation of pure-phase $CuGaO_2$ as well as the large particle sizes, disfavor its application for perovskite solar cells and other optoelectronic applications. Consequently, an alternative method for the development of $CuGaO_2$ nanostructures under mild reaction conditions which produce ultrafine nanoparticles with controllable morphology is of high importance [26].

To date, solid state reactions and vacuum-deposition have been extensively applied for the synthesis of $CuGaO_2$ [35,39-44]. However, both methods exhibit significant drawbacks. For example, the high crystallization temperature used in solid-state synthesis can lead to the production of large and aggregated particles with size over 1 μm. In addition, nanostructured materials cannot be easily obtained using vacuum deposition due to its demanding experimental conditions. Yu et al. (2012) have recently developed a more convenient and cost-effective synthesis method, using a relatively low temperature (~240 °C) process [26]. In this procedure, binary metal oxides and sodium hydroxide were used as precursors and mineralizer, respectively; however, the as-produced $CuGaO_2$ particles were on the micrometer scale. Recently, the hydrothermal synthesis of $CuGaO_2$ has been reported, whereas $CuGaO_2$ nanoplates with size ranging from ~1 μm to ~300 nm were produced by simply changing the pH of the precursor solution [45-49]. In a similar study, Yu and co-workers reported the phase formation and particle growth mechanism of delafossite $CuGaO_2$ particles under hydrothermal conditions using soluble metal salts as precursors [30]. The formation of nanoparticles (NPs) of size of ~20 nm was observed with the addition of sodium dodecyl sulfate (SDS). However, due to the low water solubility of this surfactant, the size of NPs cannot be precisely controlled and therefore the formation of particles with different sizes (ranging from nm to μm) takes place [26]. Moreover, delafossite $CuGaO_2$ nanoplates 50-100 nm wide and 10-20 nm thick, have been recently synthesized by Wang et al. (2015) and Zhang et al. (2017), using microwave-assisted hydrothermal reaction [23,34] and have been used in n-i-p perovskites solar cells [34]. The relative large sizes of all $CuGaO_2$ nanostructures, reported in the literature so far, might lead to the formation of non-compact films with high roughness and pinholes, which usually provide limitations on the performance of solution processed opto-electronic devices. Thus, the synthesis of $CuGaO_2$ NPs with unique characteristics, i.e. dimension in the nanometer range, high purity, spherical morphology and high yield percentage for high performance p-i-n perovskite solar cells remains a challenge.

Among various methods, the surfactant-assisted hydrothermal method can be characterized as a desirable process in the preparation of nanoscale materials. This synthesis method offers several important advantages over conventional hydrothermal methods, such as controlled size of NPs, low temperature growth, simplicity and low cost [50]. The critical role of surfactant is to control the size, shape and morphology of the NPs, as well as a decrease in the crystallization temperature favoring therefore, the formation of a pure phase $CuGaO_2$ nanomaterials and offering a cost-effective method for large scale productions.

In this work, we show one step synthesis of pure phase, ultrafine $CuGaO_2$ nanocrystals with average size ~5 nm and with uniform size distribution. The high quality $CuGaO_2$ nanocrystals were fabricated using a cost-effective, low temperature surfactant-assisted hydrothermal method, only after 4 hours of reaction in a stainless steel sealed autoclave. The ultra-low NPs average size enables the formation of compact films, with high electrical conductivity, optical transparency and low surface roughness. The nanoparticulate $CuGaO_2$ functional layers are used as HTLs between the ITO and $CH_3NH_3PbI_3$ in a p-i-n perovskite solar cell architecture. The corresponding perovskite solar cells exhibit excellent hole carrier selectivity, high Voc= 1.09 V and power conversions efficiency of 15.3 %, which is around 40 % higher compared to the PCE achieved in PEDOT: PSS-based solar cells.

**2. Results & discussion**

We conducted preliminary experiments with the aim to obtain $CuGaO_2$ NPs with dimension in the nanometer range, high purity and uniform morphology, which are essential properties for functional electronic applications. More specifically, the concentration of the precursors and reducing agent, the addition of surfactant, the pH of the solution and the reaction temperature and time (data not shown) were examined in detail. The optimum conditions are equal to those reported in the experimental section. Briefly, the addition of pluronic P123 ($EO_{20}PO_{70}EO_{20}$) block copolymer as surfactant was found to be beneficial for the synthesis of $CuGaO_2$ NPs, as it can decrease the surface energy and temperature of crystallization and control the size and morphology of NPs. High reaction temperature (> 260 °C) and time (more than 4 hours) were found to lead to the formation of large NPs size (> 100 nm). A similar trend was also observed with the increase in precursor's concentration, in accordance to literature data [30]. We also found that higher precursor's dosage (> 4 mmol) affects positively the growth mechanisms of nanocrystals. Regarding the pH of the solution, at high pH values (pH > 6.5) the formation of $Cu(OH)_2$ is favored, leading to increased impurities of $CuO_x$. While at acidic conditions (pH < 3), $[Ga(OH)_2(H_2O)_4]^+$ species are predominantly formed leading to crystalline γ-GaO(OH) particles. Powder X-ray diffraction (XRD) was used for the investigation of crystal structure and phase composition of the obtained $CuGaO_2$ NPs.

## 2.1. CuGaO₂ nanoparticles characterization

The wide-angle XRD pattern of $CuGaO_2$ displays several broad Bragg diffraction peaks, which can be assigned to the hexagonal structure of $CuGaO_2$ ($a_{hex}$ = $b_{hex}$ = 2.97 Å and $c_{hex}$ = 17.17 Å, JCPDS card no 41-0255), Fig. 1a. Please note that the XRD pattern does not present any other diffraction peak, suggesting that as-prepared material is single-phase $CuGaO_2$. The average crystallite size of $CuGaO_2$ calculated from Scherrer analysis of the (006) peak was found to be ~10 nm. For comparison, XRD analysis was also performed for the material prepared using a similar procedure, but without addition of surfactant. The analysis showed the formation of γ-GaO(OH) as the main product, whereas $CuGaO_2$ peaks were not detected (Fig. S1).

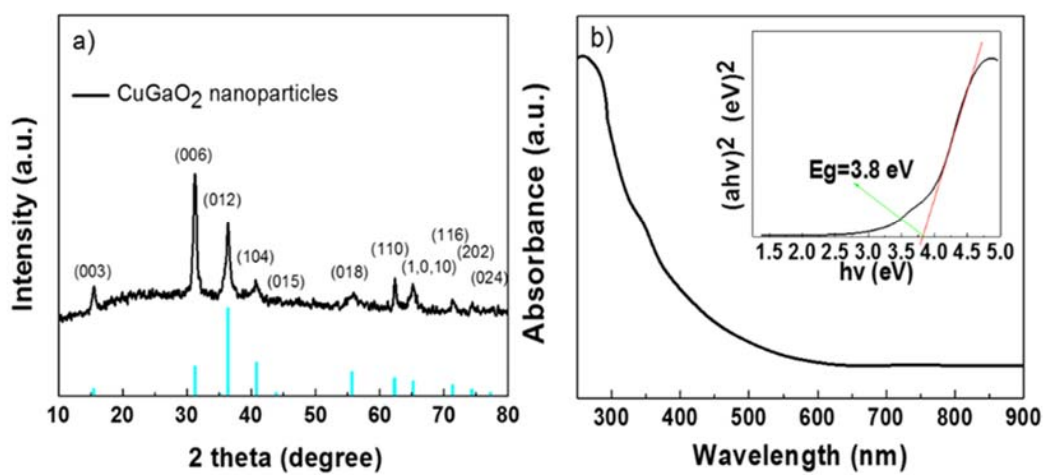

**Fig. 1.** a) Wide-angle powder XRD pattern and b) optical absorption spectrum of $CuGaO_2$ NPs. The inset in panel b shows the corresponding $(\alpha h\nu)^2$ vs. energy ($h\nu$) plot.

As shown in Fig. 1b, the optical properties of as-prepared $CuGaO_2$ NPs were characterized by diffuse reflectance ultraviolet-visible/near-IR (UV-vis/NIR) spectroscopy. The $CuGaO_2$ NPs exhibit an intense optical absorption onset in the UV region, which is associated with an optical band gap of approximately 3.8 eV (Fig. 1b). The observed wide optical band gap implies a high optical transparency of the prepared NPs in the desired wavelength range.

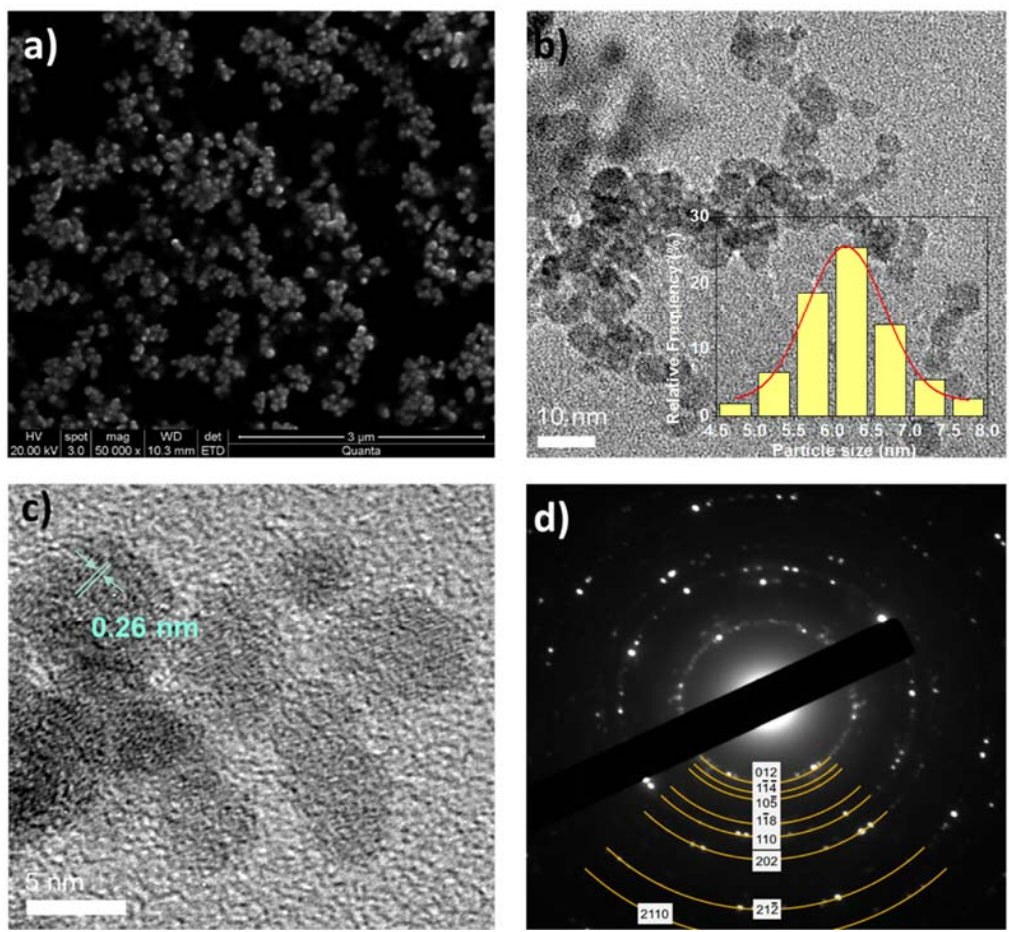

**Fig. 2.** Typical a) SEM image, b) TEM image, **inset**: Size distributions of the CuGaO$_2$ NPs obtained from TEM images c) High-resolution TEM (HRTEM) and d) SAED pattern for CuGaO$_2$ NPs.

Fig. 2 shows the advanced scanning electron microscopy (SEM) and transmission electron microscopy (TEM) measurements on the as synthesized CuGaO$_2$ NPs. It is clear that, under the present conditions, the particles do not grow into nanoplates, but rather form large agglomerates of spherical NPs. This observation suggest that our proposed surfactant-assisted hydrothermal synthesis suppresses the tendency of to form CuGaO$_2$ nanoplates [17] and promotes the formation of small-sized spherical NPs. As a comparison, SEM images of the sample prepared without surfactant points to the formation of GaO(OH) nanorods as the main structure (Fig. S2a) and with SDS as surfactant led to large nanoplates (Fig. S2b). In agreement with the SEM images, the TEM image of the CuGaO$_2$ NP aggregates (fig. 2b) clearly shows that the material is consisted of individual NPs with an average size of ~5 nm (Fig. 2b inset), which is very close to the grain size calculated from the XRD data. The high-resolution (HRTEM) image shown in Fig. 2c reveals that the constituent CuGaO$_2$ NPs possess a single-phase delafossite structure with high crystallinity; combined with XRD results, the lattice fringes with a d-spacing of ~2.6 Å can be assigned to the (110) crystal planes of delafossite CuGaO$_2$. The crystal structure of the CuGaO$_2$ NPs was further studied by selected-area electron diffraction (SAED). The SAED pattern taken from a small area of the CuGaO$_2$ NP aggregates (Fig. 2d) shows a series of spotted Debye-Scherrer diffraction rings, which can be indexed to the hexagonal phase of CuGaO$_2$ (R-3m space group). No other crystal phases were observed by means of electron diffraction. Furthermore, characterization of the chemical composition of CuGaO$_2$ NPs with energy-dispersive X-ray spectroscopy (EDS) showed an overall Cu:Ga atomic ratio close to 1:1, in agreement with the stoichiometry of CuGaO$_2$ (Fig. S3).

*2.2. Characterization of hole transport layer and perovskite film quality*

We have shown above that pure phase ultrafine CuGaO$_2$ spherical NPs are formed by following the proposed surfactant-assisted hydrothermal synthesis. The small particle sizes generally enable the formation of dense, compact films with low surface roughness, which are crucial parameters for p-i-n perovskite solar cell application. Furthermore, to avoid the agglomeration of the NPs the selection of an appropriate solvent is of utmost importance. In the present study, DMSO was used concurrently as a ligand and dispersing solvent, stabilizing the CuGaO$_2$ NPs. DMSO is an aprotic solvent with high polarity, the absence of hydrogen bonds between DMSO molecule favor the formation of colloidal NPs with minimal aggregation; this is because the oxygen and sulfur atoms in DMSO can coordinate to metal ions on nanoparticles surface [51]. In an attempt to further enhance the dispersion quality, the dispersions were treated with a high frequency sonicator probe for over 60 min and then filtered through a 0.45-μm hydrophobic polyvinylidene fluoride (PVDF) filter. The obtained colloidal solution is of excellent quality and remains stable for several months (Fig. S5). The CuGaO$_2$ NPs dispersion was spin coated on glass/ITO substrates, followed by thermal annealing at 300 °C for 20 min in ambient atmosphere resulting in the formation of a 15 nm thick film (as described in experimental section). The thermogravimetric analysis (TGA) profile of the CuGaO$_2$ nanoparticles showed a weight loses of ~2.2 % in the 105-300 °C temperature range, corresponding to the liberation of residual solvent. Then, a weight loss of about 1.7 % was observed up to 600 °C, possibly due to the dehydration and/or dihydroxylation of the CuGaO$_2$ surface (Fig. S4). As shown in Fig. 3b the CuGaO$_2$ film exhibits a high degree of compactness and smooth surface (Ra=4.2 nm), which can be attributed to the small size of CuGaO$_2$ NPs. Importantly, as shown in Fig. 3c, the CH$_3$NH$_3$PbI$_3$ active layer which is fabricated on top of ITO/ CuGaO$_2$-NPs is also quite compact and smooth (Ra=14.2 nm) consisted of relatively large grains in the range of ~300 nm. As expected from the literature the type of underlayer and processing conditions influence the morphology of the perovskite active layer [19].

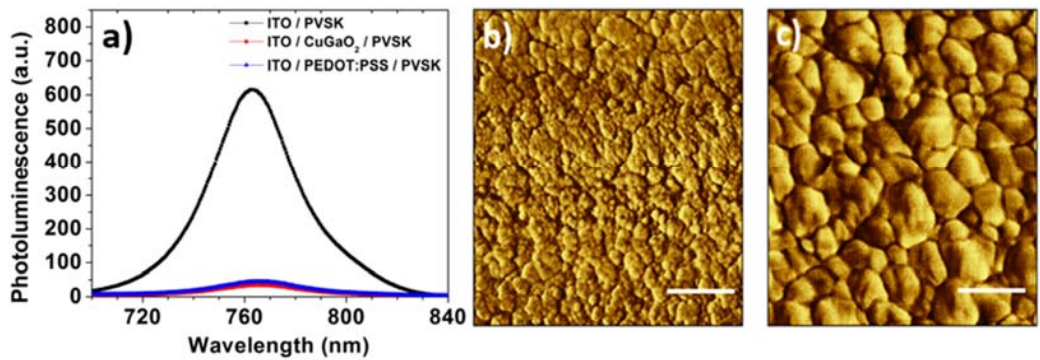

**Fig. 3.** a) Photoluminescence spectra of $CH_3NH_3PbI_3$ on top of ITO (black filled squares), ITO/$CuGaO_2$-NPs (red filled circles) and ITO/PEDOT:PSS (blue filled triangles) electrodes. AFM images of b) ITO/$CuGaO_2$ and (c) ITO/$CuGaO_2$/$CH_3NH_3PbI_3$ films. (The scale bar is 500 nm.)

Another important prerequisite for the HTLs is the carrier (hole) transport and collection efficiency of the electrode. Fig. 3a shows the steady state photoluminescence (PL) spectra of $CH_3NH_3PbI_3$ photoactive layer fabricated on top of ITO, ITO/$CuGaO_2$ and ITO/PEDOT:PSS electrodes, respectively. The PL spectrum of $CH_3NH_3PbI_3$ fabricated on top of ITO shows a strong emission at 764 nm, whereas the intensity of the corresponding peaks quenched over 95 % when the $CH_3NH_3PbI_3$ layer is fabricated on top of ITO/$CuGaO_2$ as well as ITO/PEDOT:PSS. The decrease in PL intensity coincides with a reduction in the band-to-band electron-hole recombination at $CH_3NH_3PbI_3$ film. The fact that PEDOT:PSS based device exhibits similar level of quenching but lower device performance indicates a trap-assisted non-radiative recombination process. The reason for the predominance of this type of recombination at PEDOT:PSS based device can be due to perovskite morphology on top of PEDOT:PSS underlayer having much smaller grain composition and higher density of grain boundaries [19].

Further, as depicted in Fig. 5a the transparency of both ITO/PEDOT:PSS (~50 nm) and ITO/CuGaO$_2$-NPs (~15 nm) electrodes is above 85 % in most of the visible spectrum (400-800 nm), allowing a high number of photons to reach the CH$_3$NH$_3$PbI$_3$ photoactive layer. The high transparency of ITO/CuGaO$_2$ NPs bottom electrode is of great importance for the performance of the corresponding photovoltaic devices.

These results indicate that our proposed surfactant-assisted hydrothermal synthesis of CuGaO$_2$ NPs can be successfully applied for the fabrication of high quality thin films suitable for advanced optoelectronic applications.

*2.3. Device performance*

As a final step, for evaluation of the novel synthesized CuGaO$_2$ based electronic layers, we fabricated complete p-i-n perovskite solar cells, with the structure: ITO/CuGaO$_2$-NPs/CH$_3$NH$_3$PbI$_3$/PC[70]BM/Al) as shown schematically in Fig. 4a. As reference device we fabricated a solar cell where ~15 nm CuGaO$_2$ was replaced by a ~50 nm thick PEDOT:PSS film. The choice of 50 nm PEDOT:PSS was made in order to fabricate highly reproducible reference devices. In principles, thick hole transporting layers are needed for reliable printed electronics device performance. The PEDOT:PSS conductivity is ~0.1 S/cm, which is about 10 times higher compared to CuGaO$_2$ conductivity (0.02 - 0.03 S/cm) as a result of lower metal oxide conductivity the thickness of CuGaO2 HTL is limited [52,53]. The current density versus voltage characteristic (J-V) curves of the best performing p-i-n devices under illumination and dark conditions are demonstrated in Fig. 4b and c, respectively and the photovoltaic parameters of the corresponding devices are presented in Table 1.

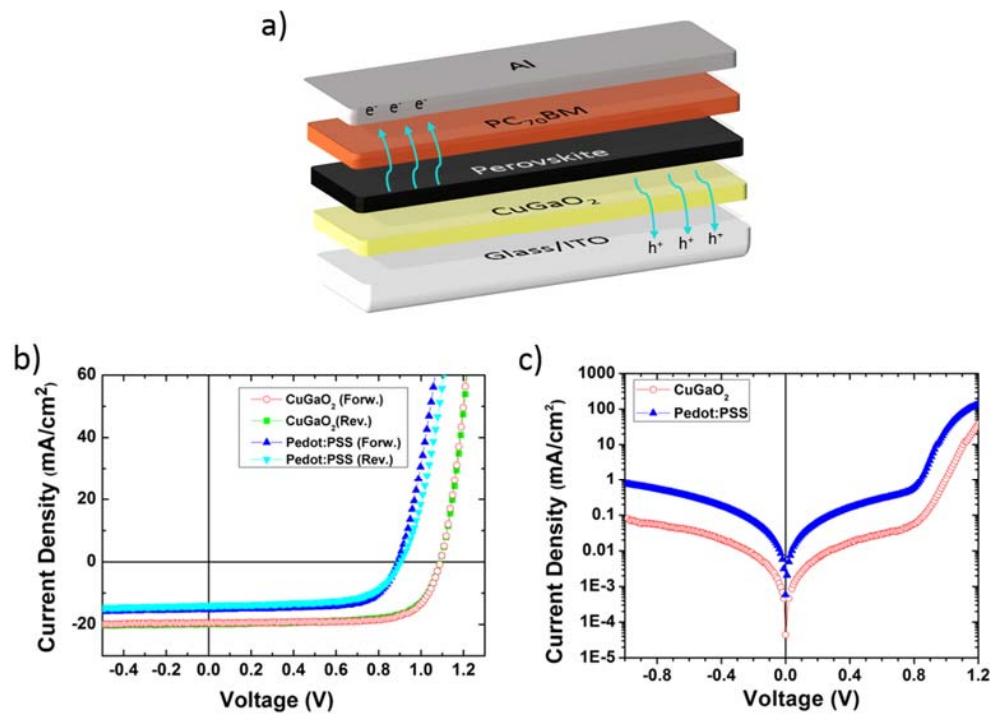

**Fig. 4.** a) The structure of the p-i-n perovskite solar cells under study (ITO/CuGaO$_2$-NPs/CH$_3$NH$_3$PbI$_3$/PC[70]BM/Al) and the current density versus voltage characteristics (J-V) plots b) under illumination, c) under dark conditions for the two p-i-n devices under this study ITO/PEDOT:PSS/CH$_3$NH$_3$PbI$_3$/PC[70]BM/Al for forward (blue filled up triangle), reverse (cyan filled down triangle) and ITO/CuGaO$_2$-NPs/CH$_3$NH$_3$PbI$_3$/PC[70]BM/Al for forward (red open circles), reverse scan direction (green filled square).

Interestingly, CuGaO$_2$ NPs based p-i-n perovskite devices exhibited a Voc of 1.09 V, which is significantly higher compared with the PEDOT:PSS-based [54]. We ascribe the enhanced Voc to the better energy level alignment of CuGaO$_2$ (~5.3 eV), with respect to perovskite's VB edge ~5.4 eV [55], compared to PEDOT:PSS (~5 eV) [56]. Therefore, hole extraction at the interface can be enhanced, inducing a lower charge recombination rate which in turn yields an increased built-in-voltage ($V_{bi}$). Furthermore, the relatively increased grain sizes of $CH_3NH_3PbI_3$ photoactive layer (Fig. 3c), and hence reduced grain boundaries area, reduces the parasitic processes that limit the FF and Voc [19,57]. It is worthy to mention that solar cell under study shows negligible hysteresis between the two different J-V scan directions (forward-reverse) (Fig. 4b) which is another important issue for solution processed hybrid perovskite solar cells. The $CuGaO_2$ NPs-based device exhibits a very good diode behavior. Specifically, as shown in Fig. 4c the leakage current under dark conditions (at Voltage = -1 V) for the $CuGaO_2$ NPs based device is 1 order of magnitude lower compared to PEDOT:PSS based device, a parameter which clearly indicates the improved $CuGaO_2$ hole carrier selectivity compared to PEDOT:PSS.

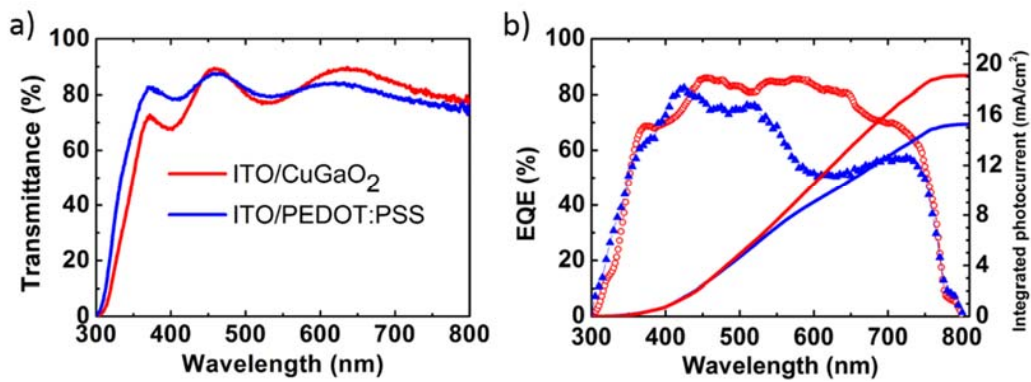

**Fig. 5.** a) Transmittance spectra of ITO/PEDOT: PSS (blue line) and ITO/$CuGaO_2$ NPs (red line), and b) EQE spectrum of p-i-n perovskite solar cells with the use of $CuGaO_2$ NPs (red cycles) and PEDOT:PSS (blue triangles) as HTL. The right axis represents the integrated photocurrent density of the corresponding perovskite solar cells.

As observed in Fig 5b, $CuGaO_2$ – based device generates more photocurrent in the spectral region between 450 and 650 nm compared with the PEDOT:PSS-based device, which declines at longer wavelengths. This behavior, is possibly due to not optimized perovskite thickness and top electrode (PC[70]BM/Al) [58]. Since both contacts show similar transparency at the specific spectral region we can ascribe the increased current density to enhance hole extraction at perovskite/$CuGaO_2$ NPs. The integrated photocurrent density (19.07 mA/cm$^2$) of $CuGaO_2$ NPs device is in good agreement with the value obtained from J-V curve (19.52 mA/cm$^2$).

**Table 1.** Extracted solar cell parameters from the characterization of the device ITO/ $CuGaO_2$/ $CH_3NH_3PbI_3$/ PC[70]BM/ Al and ITO/ PEDOT:PSS/ $CH_3NH_3PbI_3$/ PC[70]BM/ Al.

| HTL | Jsc (mA/cm$^2$) | Voc (V) | FF (%) | PCE (%) |
|---|---|---|---|---|
| *$CuGaO_2$-15nm* | 19.52 | 1.09 | 71.8 | 15.3 |
| *PEDOT:PSS-50nm* | 15.05 | 0.89 | 69.7 | 9.3 |

Importantly, metal oxides as front contact hole transporting layers can prevent the perovskite films from intense UV light exposure, and thus can be expected to improve the photostability of perovskite devices, as it has been recently reported by Zhang et al. with the use of $CuCrO_2$ as HTL [59].

To summarize, with the incorporation of the high quality dispersed $CuGaO_2$ NPs into p-i-n perovskite solar cells, an increased PCE of 15.3 % compared to that achieved with PEDOT:PSS (9.3 %) is achieved. The $CuGaO_2$ HTL based perovskite solar cells exhibited a Voc equal to 1.09 V which is much higher compared to that achieved using conventional PEDOT:PSS HTL (Voc of 0.89 V). The improved PCE is attributed to

(i) the quality of ITO/CuGaO$_2$ NPs bottom electrode as confirmed above by the efficient quenching of the photoluminescence of CH$_3$NH$_3$PbI$_3$ films, (ii) the observed enhanced crystallinity and absorption characteristics of the CH$_3$NH$_3$PbI$_3$ photoactive layer when fabricated on top of ITO/ CuGaO$_2$ NPs, (iii) the enhanced Voc due to proper energy level alignment of CuGaO$_2$, with respect to the perovskite's VB edge and (iv) due to the low leakage current density. The Fig. S6 show the statistical analysis of performance reproducibility for the corresponding p-i-n perovskite solar cells using the proposed CuGaO$_2$ hole selective contact.

## 3. Conclusions

A novel one-step synthesis of p-type delafossite CuGaO$_2$ nanocrystals by a pluronic surfactant-assistant hydrothermal method is reported. The reaction proceeds under low temperature (220 °C) and within a short reaction time (4 h). By the proposed synthetic route, the size of the NPs was fully controlled and the formation of CuGaO$_2$ NPs with diameter of ~5 nm was achieved. A detailed physicochemical characterization of the CuGaO$_2$ NPs is reported including X-ray diffraction, EDS analysis and electron microscopy studies which confirm the high purity and the small grain composition of the CuGaO$_2$ NPs. We have shown that the proposed synthetic route can be used to produce a stable dispersion of CuGaO$_2$ NPs in DMSO with high purity and crystallinity.

The optoelectronic measurements confirmed the high transparency (wide optical band-gap) and conductivity of the CuGaO$_2$ interlayer. DMSO was subsequently used as a ligand and dispersing solvent for stabilizing the CuGaO$_2$ which were successfully used as HTL in p-i-n perovskite solar cells. The corresponding 15 nm CuGaO$_2$ HTL-based

p–i–n perovskite solar cells show PCE of 15.3 %, which is significantly higher compared to the PCE values (9.3 %) obtained with conventional 50 nm PEDOT:PSS HTL-based p–i–n perovskite solar cells.

To conclude, a generic delafossite oxides synthetic route to control nanoparticle formation is proposed. It is shown that pluronic surfactant-assisted hydrothermal method can produce ultrafine (~5 nm) p-type $CuGaO_2$ nanoparticles with stable dispersion in DMSO. Functional $CuGaO_2$ hole transporting layers provide improvement on carrier selectivity and other electrode related properties for perovskite photovoltaics. We believe that the proposed surfactant-assistant hydrothermal method can provide a generic synthetic pathway to produce other delafossite oxides relevant to advanced energy materials applications.

**4. Experimental methods**

*4.1. Materials*

Pre-patterned glass-ITO substrates (sheet resistance 4Ω/sq) were purchased from Psiotec Ltd, $PbI_2$ from Alfa Aesar, MAI from Dyenamo Ltd, PC[70]BM from Solenne BV and PEDOT:PSS from (Clevios PH). All the other chemicals used in this study were purchased from sigma Aldrich.

*4.2. Synthesis of $CuGaO_2$ nanoparticles*

1.35 mmol of $Cu(NO_3)_2.6H_2O$, $Ga(NO_3)_3.6H_2O$ and 2 g Pluronic 123 surfactant was dissolved in deionized water with the pH tuned to 3.2 by the slowly addition of KOH (1M) solution with stirring. The final volume of the precursor was kept at 80 mL with 2 mL of ethylene glycol added. The 3 hours-stirred precursors was then sealed in a 100 mL stainless-steel autoclave to undergo a hydrothermal reaction at 220 °C for 4 h in a preheated oven and quenching under flowing water. After the reaction, the obtained nanoparticles with a light-yellow color were collected by centrifuging and then were successively washed with diluted aqueous ammonia solution (5 % v/v) and acetic acid (5 M) (four washes for 30 min each time) to remove any possible traces of $Cu_2O$, CuO or Cu and Ga(III) species respectively. Finally, the as-synthesized samples were washed with distilled water and ethanol for several times before characterizations. The $CuGaO_2$ NPs were stored in DMSO solution for further use. For comparison purposes, the same procedure described above was also used for the preparation of NCAs without surfactants and with SDS as surfactant.

*4.3. Device fabrication*

The inverted solar cells under study was ITO/PEDOT:PSS or $CuGaO_2$-NPs/$CH_3NH_3PbI_3$/PC[70]BM/Al. ITO substrates were sonicated in acetone and subsequently in isopropanol for 10 min and heated at 120 °C on a hot plate 10 min before use. To form 50 nm PEDOT:PSS, 35 µL ink was dynamically spin coated at 6000 rpm for 30 s in air on a preheated ITO substrate and then annealing at 120 °C for 25 minutes in a nitrogen filled glove box. To form a ~15 nm $CuGaO_2$ NPs thin film as HTL the as prepared dispersions were diluted to 5 mg ml$^{-1}$ in DMSO and dynamically spin coated at 4000 rpm/30 secs/80 ul and then thermally annealed at 300 °C for 20 min in ambient atmosphere. The $CuGaO_2$ NPs dispersion was improved using probe sonicator (Qsonica 700 w) for 60 min and 0.45 µm PVDF filters for filtration. The perovskite solution was prepared 30 min prior spin coating by mixing

Pb(CH$_3$CO$_2$)$_2$·3H$_2$O: Methylammonium iodide (1:3) at 36 %wt in DMF with the addition of 1.5 % mole of methylammonium bromide (MABr) [54]. The precursor was filtered with 0.1 µm PTFE filters. The perovskite precursor solution was deposited on the HTLs by static spin coating at 4 k rpm for 60 seconds and annealing for 5 minutes at 80 °C, resulting in a film with a thickness of ~250 nm. The PC[70]BM solution, 20 mg ml$^{-1}$ in chlorobenzene, was dynamically spin coated on the perovskite layer at 1k rpm for 30 sec. Finally, 100 nm Al layers were thermally evaporated through a shadow mask to finalize the devices. Encapsulation was applied directly after evaporation in the glove box using a glass coverslip and an Ossila E131 encapsulation epoxy resin activated by 365 nm UV-irradiation. The active area of the devices was 0.09 mm$^2$.

*4.4. Characterization*

X-ray diffraction (XRD) patterns were collected on a PANanalytical X´pert Pro MPD powder diffractometer (40 kV, 45 mA) using Cu Kα radiation (λ=1.5418 Å). Transmission electron microscope (TEM) images and electron diffraction patterns were recorded on a JEOL JEM-2100 microscope with an acceleration voltage of 200 kV. The samples were first gently ground, suspended in ethanol and then picked up on a carbon-coated Cu grid. Quantitative microprobe analyses were performed on a JEOL JSM-6390LV scanning electron microscope (SEM) equipped with an Oxford INCA PentaFET-x3 energy dispersive X-ray spectroscopy (EDS) detector. Data acquisition was performed with an accelerating voltage of 20 kV and 60 s accumulation time. Thermogravimetric analysis (TGA) on the CuGaO$_2$ nanoparticles was performed using a Shimadzu TG system. TGA was conducted from 40 to 600 °C in air atmosphere (200 mL min$^{-1}$ flow rate) with a heating rate of 5 °C min$^{-1}$. Transmittance and absorption measurements were performed with a Schimadzu UV-2700 UV-Vis spectrophotometer. Diffuse reflectance UV–vis spectra were recorded at room temperature with a Schimadzu UV-2700 UV-Vis optical spectrophotometer, using powder $BaSO_4$ as a 100 % reflectance standard. Reflectance data were converted to absorption (α/S) date according to the Kubelka-Munk equation: $α/S = (1-R)^2/(2R)$, where R is the reflectance and α, S are the absorption and scattering coefficients, respectively [60]. The thickness of the active layers was measured with a Veeco Dektak 150 profilometer. The current density-voltage (J/V) characteristics were characterized with a Botest LIV Functionality Test System. Both forward and reverse scans were measured with 10 mV voltage steps and 40 msec of delay time. For illumination, a calibrated Newport Solar simulator equipped with a Xe lamp was used, providing an AM1.5G spectrum at 100 mW/cm$^2$ as measured by a certified oriel 91150V calibration cell. A custom-made shadow mask was attached to each device prior to measurements to accurately define the corresponding device area. Atomic force microscopy (AFM) images were obtained using a Nanosurf easy scan 2 controller under the tapping mode. EQE measurements were performed by Newport System, Model 70356_70316NS.

**Acknowledgements**

We would like to thank Cyprus University of Technology for providing basic internal budget for the maintenance of the Molecular Electronics and Photonics Research unit and the European Research Council (ERC) under the European Union's Horizon 2020 research and innovation program (grant agreement No 647311) for funding.

**Conflict of interest**

The authors declare no conflict of interest.

**Appendix A. Supporting information**

Supplementary data associated with this article can be found in the online version at http://dx.doi.org/10.1016/j.mtener.xxxx.xx.xxx.

**References**

[1] N. Ahn, D.-Y. Son, I.-H. Jang, S. M, Kang, M. Choi, N.-G. Park, Highly reproducible perovskite solar cells with average efficiency of 18.3 % and best efficiency of 19.7 % fabricated via Lewis base adduct of lead(II) iodide, J. Am. Chem. Soc. 137 (2015) 8696-8699.

[2] J.-H. Im, I.-H. Jang, N. Pellet, M. Gratzel, N.-G. Park, Growth of $CH_3NH_3PbI_3$ cuboids with controlled size for high-efficiency perovskite solar cells, Nat. Nanotechnol. 9 (2014) 927-932.

[3] H. Zhou, Q. Chen, G. Li, S. Luo, T.-B. Song, H.-S. Duan, Z. Hang, J. You, Y. Liu, Y. Yang, Photovoltaics. Interface engineering of highly efficient perovskite solar cells, Science 345 (2014) 542-546.

[4] X. Gong, M. Li, X. B. Shi, H. Ma, Z. K. Wang, L. S. Liao, Controllable Perovskite Crystallization by Water Additive for High-Performance Solar Cells, Adv. Funct. Mater. 25 (2015) 6671-6678.

[5] G.-C. Xing, N. Mathews, S.Y. Sun, S. S. Lim, Y. M. Lam, M. Gratzel, S. Mhaisalkar, T. C. Sum, Long-range balanced electron- and hole-transport lengths in organic-inorganic $CH_3NH_3PbI_3$, Science 342 (2013) 344-347.

[6] N. J. Jeon, J. H. Noh, W. S. Yang, Y. C. Kim, S. Ryu, J. Seo, S. I. Seok, Compositional engineering of perovskite materials for high-performance solar cell, Nature 517 (2015) 476-480.

[7] C.-H. Chiang, Z.-L. Tseng, C.-G. Wu, Planar heterojunction perovskite/$PC_{71}BM$ solar cells with enhanced open-circuit voltage via a (2/1)-step spin-coating process, J. Mater. Chem. A 2 (2014) 15897-15903.

[8] A. Komija, K. Teshima, Y. Shirai, T. Miyasaka, Organometal halide perovskites as visible-light sensitizers for photovoltaic cells, J. Am. Chem. Soc. 131 (2009) 6050-6051.

[9] W. S. Yang, J. H. Noh, N. J. Jeon, Y. C. Kim, S. Ryu, J. Seo, S. I. Seok, SOLAR CELLS. High-performance photovoltaic perovskite layers fabricated through intramolecular exchange, Science 348 (2015) 1234-1237.

[10] Q. Chen, H. Zhou, Z. Hong, S. Luo, H.-S. Duan, H.-H. Wang, Y. Liu, G. Li, Y. Yang, Planar heterojunction perovskite solar cells via vapor-assisted solution process, J. Am. Chem. Soc. 136 (2014) 622-625.

[11] F. Igbari, M. Li, Y. Hu, Z.-K. Wang, L.-S. Liao, A room-temperature $CuAlO_2$ hole interfacial layer for efficient and stable planar perovskite solar cells, J. Mater. Chem. A 4 (2016) 1326-1335.

[12] Z. Xiao, Q. Dong, C. Bi, Y. Shao, Y. Yuan, J. Huang, Solvent annealing of perovskite-induced crystal growth for photovoltaic-device efficiency enhancement, Adv. Mater.

26 (2014) 6503-6509.

[13] J. You, Z. Hong, Y. (M.) Yang, Q. Chen, M. Cai, T.-B. Song, C.-C. Chen, S. Lu, Y. Liu, H. Zhou and Y. Yang, Low-temperature solution-processed perovskite solar cells with high efficiency and flexibility, ACS Nano 8 (2014) 1674-1680.

[14] N. Tripathi, M. Yanagida, Y. Shirai, T. Masuda, L. Han, K. Miyano, Hysteresis-free and highly stable perovskite solar cells produced via a chlorine-mediated interdiffusion method, J. Mater. Chem. A 3 (2015) 12081-12088.

[15] Q. Xue, Z. Hu, J. Liu, J. Lin, C. Sun, Z. Chen, C. Duan, J. Wang, C. Liao, W. M. Lau, F. Huang, H.-L. Yip, Y. Cao, Highly efficient fullerene/perovskite planar heterojunction solar cells *via* cathode modification with an amino-functionalized polymer interlayer, J. Mater. Chem. A 2 (2014) 19598-19603.

[16] J. H. Kim, P.-W. Liang, S. T. Williams, N. Cho, C.-C. Chueh, M. S. Glaz, D. S. Ginger, A. K.-Y. Jen, High-performance and environmentally stable planar heterojunction perovskite solar cells based on a solution-processed copper-doped nickel oxide hole-transporting layer, Adv. Mater. 27 (2015) 695-701.

[17] Z. K. Wang, M. Li, D. X. Yuan, X. B. Shi, H. Ma, L. S. Liao, Improved hole interfacial layer for planar perovskite solar cells with efficiency exceeding 15%, ACS Appl. Mater. Interf. 7 (2015) 9645-9651.

[18] K. X. Steirer, P. F. Ndione, N. E. Widjonarko, M. T. Lloyd, J. Meyer, E. Ratcliff L., A. Kahn, N. R. Armstrong, C. J. Curtis, D. S. Ginley, Enhanced efficiency in plastic solar cells via energy matched solution processed NiOx interlayers, Adv. Energy Mater. 1 (2011) 813-820.

[19] F. Galatopoulos, A. Savva, I. T. Papadas, S. A. Choulis, The effect of hole transporting layer in charge accumulation properties of p-i-n perovskite solar cells, Applied Physics Letters 5 (2017) 76102-7.

[20] J. R. Manders, S.-W. Tsang, M. J. Hartel, T.-H. Lai, S. Chen, C. M. Amb, J. R. Reynolds, F. So, Solution-Processed Nickel Oxide Hole Transport Layers in High Efficiency Polymer Photovoltaic Cells, Adv. Funct. Mater. 23 (2013) 2993-3001.

[21] V. Shrotriya, G. Li, Y. Yao, C.-W. Chu, Y. Yang, Transition metal oxides as the buffer layer for polymer photovoltaic cells, Appl. Phys. Lett. 88 (2006) 73508-3.

[22] M. T. Greiner, M. G. Helander, W.-M. Tang, Z.-B. Wang, J. Qiu, Z.-H. Lu, Universal energy-level alignment of molecules on metal oxides, Nat. Mater. 11 (2012) 76-81.

[23] J. Wang, V. Ibarra, D, Barrera, L. Xu, Y.-J. Lee, J. W. P. Hsu., Solution synthesized p-type copper gallium oxide nanoplates as hole transport layer for organic photovoltaic devices, J. Phys. Chem. Lett. 6 (2015) 1071-1075.

[24] M. M.-Masis, S. De Wolf, R. W.-Robinson, J. W. Ager, C. Ballif, Transparent Electrodes for Efficient Optoelectronics, Adv. Electron. Mater. 3 (2017) 1600529.

[25] W. A. D.-Shohl, T. B. Daunis, X. Wang, J. Wang, B. Zhang, D. Barrera, Y. Yan, J. W. P. Hsu, D. B. Mitzi, Room-temperature fabrication of a delafossite $CuCrO_2$ hole transport layer for perovskite solar cells, J. Mater. Chem. A 6 (2018) 469-477.

[26] M. Yu, T.I. Draskovic, Y. Wu, Understanding the crystallization mechanism of delafossite $CuGaO_2$ for controlled hydrothermal synthesis of nanoparticles and nanoplates, Inorg. Chem. 53 (2014) 5845-5851.

[27] S. Kato, R. Fujimaki, M. Ogasawara, T. Wakabayashi; Y. Nakahara, S. Nakata,

Oxygen storage capacity of CuMO$_2$ (M = Al, Fe, Mn, Ga) with a delafossite-type structure, Appl. Catal. B 89 (2009) 183-188.

[28] J. W. Lekse, M. K. Underwood, J. P. Lewis, C. J. Matranga, Synthesis, characterization, electronic structure, and photocatalytic behavior of CuGaO$_2$ and CuGa$_{1-x}$Fe$_x$O$_2$ (x = 0.05, 0.10, 0.15, 0.20) delafossites, Phys. Chem. C 116 (2012) 1865-1872.

[29] M. Yu, G. Natu, Z. Ji, Y. Wu, p-Type dye-sensitized solar cells based on delafossite CuGaO$_2$ nanoplates with saturation photovoltages exceeding 460 mV, J. Phys. Chem. Lett. 3 (2012) 1074-1078.

[30] M. Yu, T. I. Draskovic, Y. Wu, Cu(I)-based delafossite compounds as photocathodes in p-type dye-sensitized solar cells, Phys. Chem. Chem. Phys. 16 (2014) 5026-5033.

[31] I. Herraiz-Cardona, F. Fabregat-Santiago, A. Renaud, B. Julián- López, F. Odobel, L. Cario, S. Jobic, S Giménez, Hole conductivity and acceptor density of p-type CuGaO$_2$ nanoparticles determined by impedance spectroscopy: The effect of Mg doping, Electrochim. Acta 113 (2013) 570-574.

[32] Z. Xu, D. Xiong, H. Wang, W. Zhang, X. Zeng, L. Ming, W. Chen, X. Xu, J. Cui, M. Wang, S. Powar, U. Bach, Y.-B. Cheng, Remarkable photocurrent of p-type dye-sensitized solar cell achieved by size controlled CuGaO$_2$ nanoplates, J. Mater. Chem. A 2 (2014) 2968-2976.

[33] D. Xiong, Q. Zhang, S. Kumar Verma, H. Li, W. Chen, X. Zhao., Use of delafossite oxides CuCr$_{1-x}$Ga$_x$O$_2$ nanocrystals in p-type dye-sensitized solar cell, J. Alloys and

Compounds 602 (2016) 374-380.

[34] H. Zhang, H. Wang, W. Chein, A.K.-Y. Jen, CuGaO$_2$: A promising inorganic hole-transporting material for highly efficient and stable perovskite solar cells, Adv. Mater. 29 (2017) 1604984-8.

[35] F. A. Benko, F. P. Koffyberg, The optical interband transitions of the semiconductor CuGaO$_2$, Phys. Status Solidi A 94 (1986) 231-234.

[36] X. Nie, S.-H. Wei, S. B. Zhang, Bipolar doping and band-gap anomalies in delafossite transparent conductive oxides, Phys. Rev. Lett. 88 (2002) 066405-4.

[37] Z.-J. Fang, C. Fang, L.-J. Shi, Y.-H. Liu, M.-C. He, First-principles study of defects in CuGaO$_2$, Chin. Phys. Lett. 25 (2008) 2997-3000.

[38] R. Gillen, J. Robertson, Band structure calculations of CuAlO$_2$, CuGaO$_2$, CuInO$_2$, and CuCrO$_2$ by screened exchange, J. Phys. Rev. B, 84 (2011) 035125-7.

[39] S. Park, D. A. Keszler, Synthesis of 3R-CuMO$_{2+\delta}$ (M=Ga, Sc, In), J. Solid State Chem.173 (2003) 355-358.

[40] R. B. Gall, N. Ashmore, M. A. Marquardt, X. Tan, D. P. Cann, Synthesis, microstructure, and electrical properties of the delafossite compound CuGaO$_2$, J. Alloys Compd. 391 (2005) 262-266.

[41] M. J. Han, K. Jiang, J. Z. Zhang, Y. W. Li, Z. G. Hu, J. H. Chu, Temperature dependent phonon evolutions and optical properties of highly c-axis oriented CuGaO$_2$ semiconductor films grown by the sol-gel method, Appl. Phys. Lett. 99 (2011) 131104-3.

[42] K. Ueda, T. Hase, H. Yanagi, H. Kawazoe, H. Hosono, H. Ohta, M. Orita, M. J. Hirano, Epitaxial growth of transparent p-type conducting $CuGaO_2$ thin films on sapphire (001) substrates by pulsed laser deposition, Appl. Phys. 89 (2001) 1790-1793.

[43] R. B. Gall, N. Ashmore, M. A. Marquardt, X. Tan, D. P. Cann, Synthesis, microstructure, and electrical properties of the delafossite compound $CuGaO_2$, *J. Alloys Compd.* 391 (2005) 262-266.

[44] V. Varadarajan, D. P. Norton, $CuGaO_2$ thin film synthesis using hydrogen-assisted pulsed laser deposition, Appl. Phys. A: Mater. Sci. Process. 85 (2006) 117-120.

[45] S. Kumar, M. Miclau, C. Martin, Hydrothermal Synthesis of $AgCrO_2$ delafossite in supercritical water: A new single-step process, Chem. Mater. 25 (2013) 2083-2088.

[46] R. Srinivasan, B. Chavillon, C. Doussier-Brochard, L. Cario, M. Paris, E. Gautron, P. Deniard, F. Odobel, S. Jobic, Tuning the size and color of the p-type wide band gap delafossite semiconductor $CuGaO_2$ with ethylene glycol assisted hydrothermal synthesis, J. Mater. Chem. 18 (2008) 5647-5653.

[47] W. C. Sheets, E. Mugnier, A. Barnabe, T. J. Marks, K. R. Poeppelmeier, Hydrothermal Synthesis of Delafossite-Type Oxides, Chem. Mater. 18 (2006) 7-20.

[48] D. Y. Shahriari, A. Barnabe, T. O. Mason, K. R. Poeppelmeier, A high-yield hydrothermal preparation of $CuAlO_2$, Inorg. Chem. 40 (2001) 5734-5735.

[49] B. Chavillon, L. Cario, C. Doussier-Brochard, R. Srinivasan, L. Le Pleux, Y. Pellegrin, E. Blart, F. Odobel, S. Jobic, Synthesis of light-coloured nanoparticles of wide band gap p-type semiconductors $CuGaO_2$ and LaOCuS by low temperature hydro/solvothermal processes, Phys. Status Solidi A 207 (2010) 1642-1646.

[50] S. Thirumalairajan, V.R. Mastelaro, A novel organic pollutants gas sensing material p-type $CuAlO_2$ microsphere constituted of nanoparticles for environmental remediation, Sensors and Actuators B 223 (2016) 138-148.

[51] K. Xu, Q. Han, DMSO as a solvent/ligand to monodisperse CdS spherical nanoparticles, J. Nanopart. Res. 18:16 (2016) 1-10.

[52] N. A. Ashmore, D. P. Cann, Electrical and structural characteristics of non-stoichiometric Cu-based delafossites, J. Mat. Sci. 40 (2005) 3891-3896.

[53] R. S. Yu, Y. C. Lee, Effects of annealing on the optical and electrical properties of sputter-deposited $CuGaO_2$ thin films, Thin Solid Films 646 (2018) 143-149.

[54] L. Zhao, D. Luo, J. Wu, Q. Hu, W. Zhang, K. Chen, T. Liu, Y. Liu, Y. Zhang, F. Liu, T. P. Russell, H. J. Snaith, R. Zhu, Q. Gong, High-performance inverted planar heterojunction perovskite solar cells based on lead acetate precursor with efficiency exceeding 18 %, Adv. Funct. Mater. 26 (2016) 3508-3514.

[55] K. G. Lim , H. B. Kim , J. Jeong , H. Kim , J. Y. Kim , T. W. Lee, Boosting the power conversion efficiency of perovskite solar cells using self-organized polymeric hole extraction layers with high work function, Adv. Mater. 26 (2014) 6461-6466.

[56] A.M. Nardes, M. Kemerink, M.M. de Kok, E. Vinken, K. Maturova, R.A.J. Janssen, Conductivity, work function, and environmental stability of PEDOT:PSS thin films treated with sorbitol, Org. Electron. 9 (2008) 727-734.

[57] S. Bag, M. F. Durstock, Large perovskite grain growth in low-temperature solution-processed planar p-i-n solar cells by sodium addition, ACS Appl. Mater. Interf. 8 (2016) 5053-5057.

[58] A. Savva, I. Burgués-Ceballos, S. A. Choulis, Improved performance and reliability of p-i-n perovskite solar cells via doped metal oxides, Adv. Energy Mater. 6 (2016) 1600285-8.

[59] H. Zhang, H. Wang, H. Zhu, C.-C. Chueh, W. Chen, S. Yang, A.K.-Y. Jen, Low-temperature solution-processed $CuCrO_2$ hole-transporting layer for efficient and photostable perovskite solar cells, Adv. Energy Mater. (2018) 1702762-9.

[60] P. Kubelka, F. Munk, Ein beitrag zur optik der farbanstriche, Tech. Z. Phys 12 (1931) 593-601.

# Supporting Information

**Employing surfactant-assisted hydrothermal synthesis to control $CuGaO_2$ nanoparticle formation and improved carrier selectivity of perovskite solar cells**

Ioannis T. Papadas[a], Achilleas Savva[a], Apostolos Ioakeimidis[a], Polyvios Eleftheriou[a], Gerasimos S. Armatas[b] and Stelios A. Choulis[a]*

[a] *Molecular Electronics and Photonics Research Unit, Department of Mechanical Engineering and Materials Science and Engineering, Cyprus University of Technology, Limassol, Cyprus.*

[b] *Department of Materials Science and Technology, University of Crete, Heraklion 71003, Greece.*

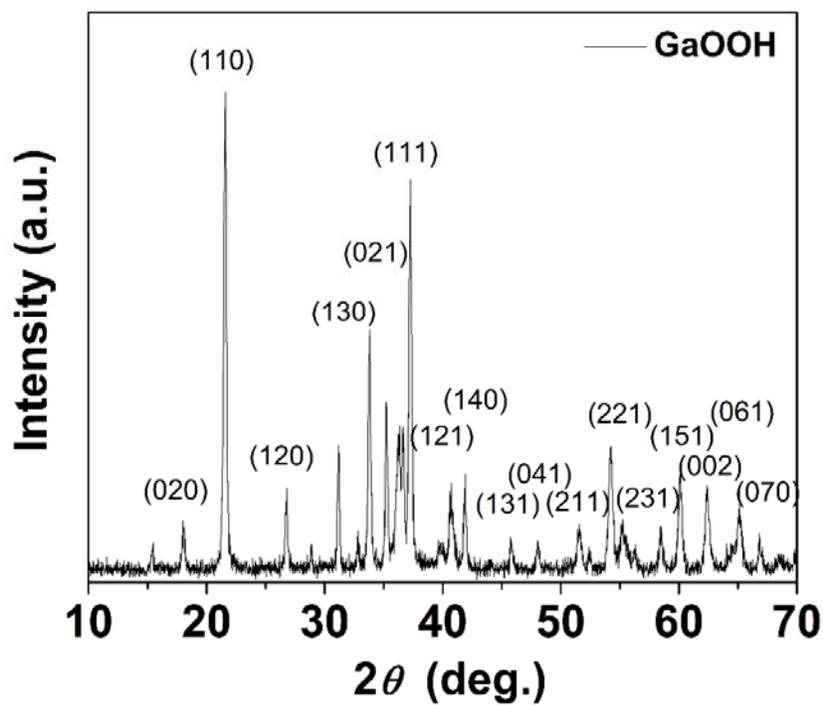

**Figure S1.** XRD pattern for the material prepared without surfactant.

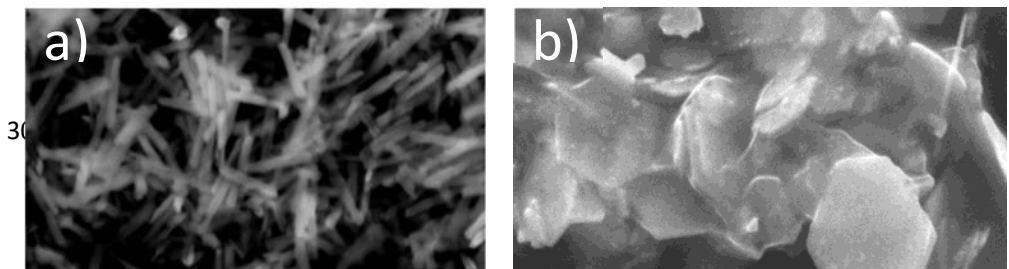

**Figure S2.** SEM image of the material produced a) without surfactant and b) with SDS surfactant.

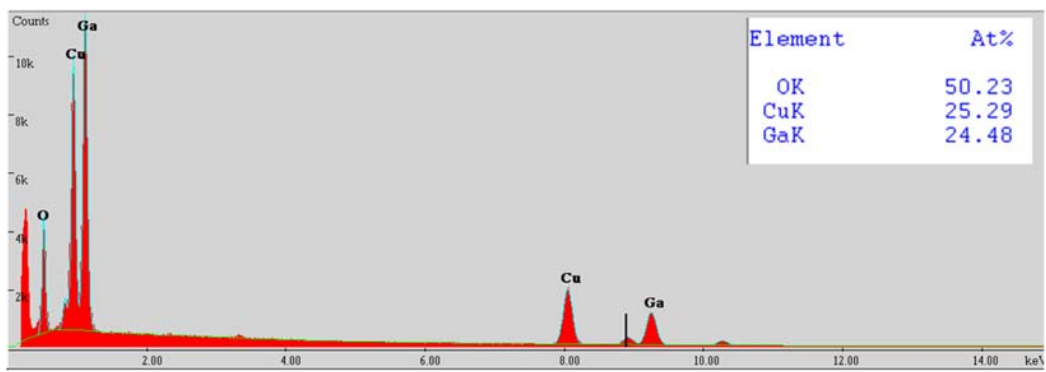

**Figure S3.** Typical EDS spectrum for CuGaO$_2$ NPs. The EDS analysis indicates an average atomic proportion of Cu:Ga ~1:1.

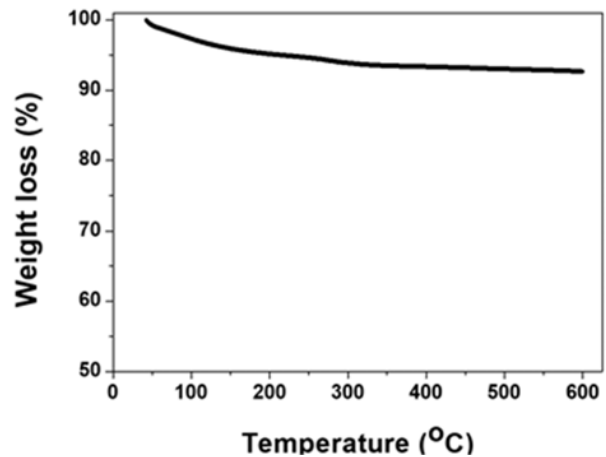

**Figure S4.** TGA profiles for as-prepared CuGaO$_2$ nanoparticles (black solid line) recorded under air flow.

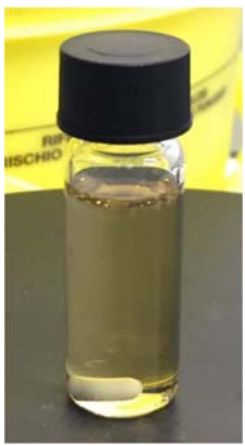

**Figure S5**. Dispersion of CuGaO$_2$ NPs in DMSO.

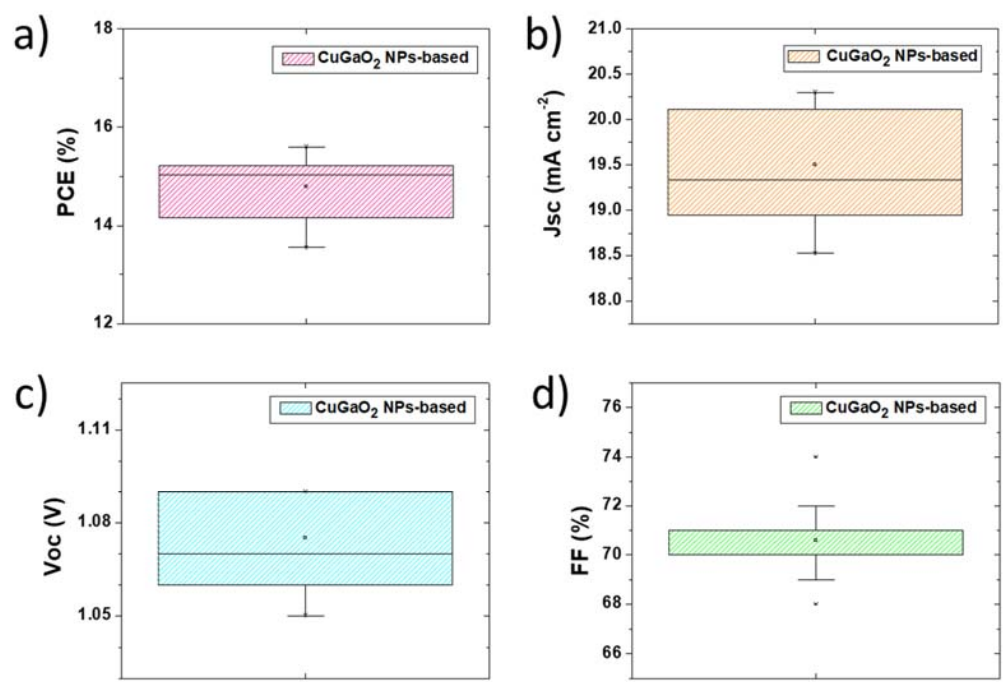

**Figure. S6:** Average photovoltaic parameters represented in box plots out of 12 devices of each series of p-i-n perovskite solar cells under study. CuGaO$_2$ NPs-based

devices with box plots, a) power conversion efficiency (PCE), b) current density (Jsc), c) open circuit voltage (Voc) and d) fill factor (FF).